\documentclass[a4paper,11pt]{article}
\pdfoutput=1 

\usepackage{jheppub} 

\usepackage[T1]{fontenc} 

\usepackage{bbold}
\newcommand{\cC}{{\cal C}}  \newcommand{\cD}{{\cal D}}

\newcommand{\cI}{{\cal I}}  
  
  \newcommand{\cN}{{\cal N}}
\newcommand{\cO}{{\cal O}}  \newcommand{\cP}{{\cal P}}

\newcommand{\cW}{{\cal W}}

\newcommand{\be}{\begin{equation}} \newcommand{\ee}{\end{equation}}
\newcommand{\bea}{\begin{eqnarray}} \newcommand{\eea}{\end{eqnarray}}
\newcommand{\beann}{\begin{eqnarray*}}  \newcommand{\eeann}{\end{eqnarray*}}
\newcommand{\bfig}{\begin{figure}} \newcommand{\efig}{\end{figure}}
\newcommand{\ba}{\begin{array}} \newcommand{\ea}{\end{array}}
\newcommand{\bcen}{\begin{center}} \newcommand{\ecen}{\end{center}}
\newcommand{\btab}{\begin{tabular}} \newcommand{\etab}{\end{tabular}}

\def\tr{\operatorname{tr\,}}

     \def\sign{\operatorname{sign}}
   
\newcommand{\bra}[1]{\langle #1|}
\newcommand{\ket}[1]{|#1\rangle}
\newcommand{\vev}[1]{\left\langle{#1}\right\rangle}

\newtheorem{Proposition}{Proposition}[section]

\newtheorem{Theorem}{Theorem}[section]
\newtheorem{Lemma}{Lemma}[section]

\newcommand{\bp}{\begin{Proposition}}   \newcommand{\ep}{\end{Proposition}}
\newcommand{\bt}{\begin{Theorem}}   \newcommand{\et}{\end{Theorem}}
\newcommand{\bl}{\begin{Lemma}}     \newcommand{\el}{\end{Lemma}}
\newcommand{\bc}{\begin{Corolary}} \newcommand{\ec}{\end{Corolary}}
\def\bra{\left\langle}
\def\ket{\right\rangle}
\def\vev#1{{\bra #1 \ket}}

\title{A defect action for Wilson loops}

 \author{Carlos Hoyos}
\affiliation{Department of Physics, Universidad de Oviedo\\
c/ Federico Garcia Lorca 18, ES-33007 Oviedo, Spain}
\emailAdd{hoyoscarlos@uniovi.es}

\abstract{An effective action is proposed to compute the expectation value of Wilson loops in $(S)U(N)$ gauge theories. The action consists of fermions localized on the loop and an Abelian gauge field that fixes the representation. The discussion is limited to weak coupling and Wilson loops in the fundamental representation extended along a smooth curve, but there are no restrictions on the matter content as long as the theory has a UV fixed point or it is conformal. For a circular Wilson loop it is found that the expectation value coincides at leading order with the exact result of the $1/2$ BPS Wilson loop of $\cN=4$ super Yang-Mills, which is determined by a solvable Gaussian matrix model. This hints towards a universal connection to string theory duals and SYK models.}

\begin{document} 
\maketitle
\flushbottom

\section{Introduction}

Wilson loop operators are some of the most fundamental observables in gauge theories. They can be defined on any closed curved $\cC$ as the path-ordered exponential of the holonomy of the gauge field on the curve
\be
\cW(\cC)=\tr\, \cP\,e^{i\oint_\cC A},
\ee
with the trace in the fundamental representation. In pure Yang-Mills they, and their products, form a complete basis of gauge-invariant observables. In general, computing the expectation value of a Wilson loop is an arduous task even at weak coupling, as it requires summing over an infinite number of diagrams even when interactions are neglected.

Remarkably, in supersymmetric theories there are closely related BPS line operators\footnote{For general curves the operators look BPS along small segments, but supersymmetry is broken. In a slight abuse of language BPS will be used in the local sense.} whose expectation value takes a very simple form for symmetric shapes, like a straight line or a circle. The simplest example is the $1/2$ BPS loop of $\cN=4$ super Yang-Mills (SYM)
\be\label{eq:bpsW}
\cW_{BPS}(\cC)=\tr\, \cP\,e^{i\oint_\cC (A_\mu dx^\mu+ \phi^I\theta_I ds)},
\ee
where $\phi^I$ are the six scalar fields of $\cN=4$ SYM and $\theta_I$ ( $\theta^2=1$) are parameters associated to the $SO(6)_R$ symmetry. The expectation value of the BPS operator on a line is a trivial constant (independent of the coupling) or it is determined by a Gaussian matrix model \cite{Erickson:2000af,Drukker:2000rr,Pestun:2007rz}.  That the two are different could be surprising at first sight, as $\cN=4$ SYM is conformal invariant and the line and the circle are related by a conformal transformation. The origin of the difference can be understood in terms of boundary conditions \cite{Erickson:2000af}  or a conformal anomaly \cite{Drukker:2000rr}. Nevertheless, conformal invariance implies that the expectation value of the Wilson loop is independent of the length of the circle.

The expectation value of the circular BPS loop in $\cN=4$ $U(N)$ SYM was first computed in the large-$N$, strong 't Hooft coupling limit by means of the AdS/CFT correspondence \cite{Berenstein:1998ij,Drukker:1999zq}, following \cite{Maldacena:1998im,Rey:1998ik}.
At weak coupling, Erickson, Semenoff and Zarembo \cite{Erickson:2000af} found the leading behavior using a resummation of ladder planar diagrams. They showed that the planar expectation value is determined by a large-$N$ Gaussian matrix model, and they speculated that diagrams with internal vertices would not modify the result, as the extrapolation to strong coupling agreed with the result from AdS/CFT. Soon after, Drukker and Gross \cite{Drukker:2000rr}  argued that the relation to the matrix model extends beyond the planar limit, and that the matrix model determines the exact result to all orders in the coupling and $N$, although the question of whether interactions modify the matrix model was still open. That the Gaussian matrix model determines the expectation value was finally established by Pestun \cite{Pestun:2007rz}, who derived the exact result using supersymmetric localization for a circular BPS loop on the sphere. The result from localization also showed that for less symmetric $\cN=2$ SYM theories, perturbative and non-perturbative corrections do appear, with the perturbative ones limited to one-loop.

These very appealing results suggest that a simpler description of the (non-BPS) Wilson loop may also exist. The string theory description of BPS loops hints as to what could be a frutiful approach in this direction. The holographic description of BPS loops in different representations consists of D-branes wrapping some of the directions in the internal geometry \cite{Gomis:2006sb}, either D5 branes wrapping a $S^5$ or D3 branes wrapping a $S^3$. The weak coupling description is a one-dimensional intersection of color D3 branes with the other D3 or D5 branes. On this intersection live dynamical fields corresponding to the strings connecting the color D3 branes with the other D-branes, with the matter content determined by the representation of the Wilson loop. Although supersymmetry plays a very important role regarding the properties of the BPS loop, it does not really enter in its formulation as a defect theory.

It seems reasonable to expect that a similar defect theory description applies to ordinary Wilson loops, and indeed it will be shown that such a formulation is possible and that it leads to a Gaussian matrix model for the expectation value of the circular Wilson loop to leading order in the weak coupling expansion. The leading order is not just the first correction proportional to the coupling, but it captures all ladder diagrams (planar and non-planar) and loop contributions to the gauge coupling renormalization. Furthermore, the calculation of the leading order contribution to the circular Wilson loop using the defect theory is extremely simple and only requires straightforward manipulations of matrix integrals and the calculation of a functional determinant of free fermions.  

The matrix model of the Wilson loop is the same as the one found for the $\cN=4$ SYM BPS loop. This may sound strange at first for two reasons, the first being that in the localization result of \cite{Pestun:2007rz} the matrix integral is obtained from an integration over constant values of scalar fields. However, this may be a feature which depends on the particular $Q$-exact functional chosen to implement the calculation, and indeed the arguments presented in the perturbative calculation of \cite{Erickson:2000af,Drukker:2000rr} point to a seemingly different origin of the matrix model. This leads to the second apparent issue, in the perturbative calculation of the BPS loop it was crucial that there is an exact cancellation between the gauge and scalar contributions such that the result is finite and independent of the length of the circle. This was understood as an effective reduction to a zero-dimensional theory (the matrix model). In the Wilson loop there are no contributions from the scalar and the leading order term has a linear divergence proportional to the length of the circle, so one would have expected that the Wilson loop is described by a one-dimensional rather than zero-dimensional theory, even at leading order in the coupling. This argument however relies on a particular gauge choice. A linear divergence also exists in the BPS loop, but it is gauge dependent and can be removed by going to Feynman gauge. Therefore, the linear divergence can be understood as a gauge artifact with no physical consequences for the BPS loop. It turns out that the linear divergence of the Wilson loop can also be removed by going to Yennie gauge and therefore the same type of arguments should apply to both, at least before interactions are taken into account. At leading order in the weak coupling expansion the only dependence of the expectation value on the length of the circle enters indirectly, though the renormalization of the gauge coupling.

The paper is organized as follows. In \S~\ref{sec:defect} the defect theory for $(S)U(N)$  gauge theories is introduced and it is shown that it gives the expectation value of a Wilson loop on a smooth closed curve. In \S~\ref{sec:weak} the action of the defect theory is expanded to next-to-leading order in the coupling and it is found to be free of divergences in Yennie gauge. The connection to IR divergences is also discussed. In \S~\ref{sec:circleW} the expectation value of a circular Wilson loop is computed and found to be equal to a Gaussian matrix integral at leading order in the weak coupling expansion. The dependence on the length of the circle through the renormalized gauge coupling is briefly discussed. A summary and a discussion of the results can be found in \S~\ref{sec:conc}. Some technical details have been collected in the appendices.

\section{Defect theory description of a Wilson loop}\label{sec:defect}

The discussion will apply to four-dimensional Yang-Mills theories with gauge fields in the adjoint representation of $U(N)$ or $SU(N)$
\be
A_\mu=A_\mu^a T^a,\ \  \tr(T^a T^b)=\frac{1}{2}\delta^{ab},\ \ a,b,=0,1,\dots,N^2-1.
\ee
The generators $a=1,\dots,N^2-1$ correspond to the $\mathfrak{su}(N)$ algebra, and the generator for the Abelian component of $\mathfrak{u}(N)\simeq \mathfrak{su}(N)\oplus \mathfrak{u}(1)$ is
\be
T^0=\frac{1}{\sqrt{2N}}\mathbb{1}.
\ee
The matter content can in principle be arbitrary, but it will be assumed that the theory is either asymptotically free or conformal. 

The expectation value of the Wilson loop along a closed curve $\cC$ is given by the path integral 
\be
\vev{\cW(\cC)}=\frac{1}{Z}\int D A_\mu D \Phi \, \cW(\cC)\, e^{iS_{YM}[A_\mu,\Phi]},
\ee
where $\Phi$ denotes any matter fields and $Z$ the path integral without the Wilson loop insertion. The action includes the pure Yang-Mills term plus the action of the matter fields and their interactions
\be
S_{YM}=-\int d^4 x\, \frac{1}{2g^2}\tr\left( F_{\mu\nu} F^{\mu\nu}\right)+S_{\text{matter}}[A_\mu,\Phi]
\ee
The Yang-Mills path integral will be left implicit and we will refer to the expectation value of the Wilson loop simply as ``Wilson loop'' in the rest of the discussion.

The curve $\cC$ will be parametrized by a trajectory $x(\tau)$, where the worldline is a circle defined in the range $ \tau\in[0,1)$. The trajectory closes as one completes a period along the circle $x(0)=x(1)$. The holonomy is the integral of the pullback of the gauge field on the circle $A_\tau(\tau)$:
\be
\oint_\cC A=\oint_\cC dx^\mu A_\mu=\int_0^1 d\tau\, \dot{x}^\mu A_\mu[x(\tau)]\equiv \int_0^1 d\tau\, A_\tau(\tau).
\ee
In the following it will be shown explicitly that the Wilson loop can be computed from a path integral involving dynamical fields localized on the curve $\cC$, and coupled to the Yang-Mills gauge fields. The defect fields consist of complex fermions $\chi$ in the fundamental representation of $(S)U(N)$, and a compact $U(1)$ gauge field $a_\tau$. Fermions satisfy anti-periodic boundary conditions around the circle $\chi(1)=-\chi(0)$, while the gauge field is periodic $a_\tau(1)=a_\tau(0)$. The Wilson loop is
\be
\vev{\cW(\cC)}=\cN\vev{\int \cD \chi\cD \chi^\dagger \cD a_\tau\; e^{i S_W+iS_{CS}}},
\ee
where $\cN$ is a normalization factor and the different contributions to the defect action are
\be
\begin{split}
S_W=&\int d\tau \chi^\dagger \,i(\partial_\tau-ia_\tau-i A_\tau)\chi, \\
S_{CS} =&\int d\tau \left( k\, a_\tau +\frac{1}{2}\tr A_\tau\right), \ k=\frac{N}{2}-1.
\end{split}
\ee
The coefficient $k$ of the Chen-Simons term for $a_\tau$ is chosen in such a way that integrating over the defect gauge field will fix the representation of the Wilson loop to be in the fundamental. Na\"{\i}vely, the right choice would have been $k=-1$, however it is necessary to shift by $N/2$ in order to compensate an anomalous contribution coming from the integration over the fermionic fields. For the same reason, it is necessary to add a Chern-Simons term for the Abelian component of the $U(N)$ Yang-Mills group. If the group is $SU(N)$ this term simply vanishes.
 
 \subsection{Gauge-fixing of the defect gauge field}
 
The Abelian gauge symmetry at the defect will be fixed to the Lorenz gauge in one dimension 
\be
\partial_\tau a_\tau=0.
\ee
The gauge fixing can be done following the usual BRST procedure. The gauge fixed path integral becomes
\be
\vev{\cW(\cC)}=\cN\int \cD \chi\cD \chi^\dagger \cD a_\tau \cD b\cD c\cD\bar{c} \;\vev{e^{i S_W+iS_{CS}+iS_{bc}}},
\ee
where $b$ is the Nakanishi-Lautrup field and $c$, $\bar{c}$ are the ghost and anti-ghost fields. The BRST-exact gauge-fixing action is
\be
S_{bc}=\int_0^1d\tau\,\frac{1}{e} \left( b \partial_\tau a_\tau -i\bar{c}\,\partial_\tau^2 c\right).
\ee
A factor of the einbein $e=|\dot{x}|$ is introduced to preserve invariance under reparametrizations of $\tau$. In this simple case of Abelian symmetry the ghosts decouple, so integrating them out only contributes to the total normalization. After integrating out $b$, $a_\tau$ can just be replaced by a constant $a_\tau=a_0$ and the functional integral becomes an ordinary integral over a periodic variable (as the $U(1)$ is compact).  The range of integration can be deduced from the periodicity of the Abelian holonomy. The defect theory contains a set of Abelian Wilson loop operators
\be
W_n=e^{in\int_0^1 d\tau a_\tau}, \ n\in \mathbb{Z}.
\ee
They remain invariant under large gauge transformations, which are a symmetry of the theory
\be
a_\tau\to a_\tau+\partial_\tau \lambda, \ \ \lambda =2\pi k\tau, \ \ k\in\mathbb{Z}.
\ee
Therefore, the periodicity of the Abelian holonomy is
\be
a_0\sim a_0+2\pi.
\ee
Then, the gauge-fixed path integral reduces to
\be
\vev{\cW(\cC)}=\cN \int_0^{2\pi} da_0\, \int \cD \chi\cD \chi^\dagger \;\vev{e^{i S_W[a_0]+i k a_0 +iS_{CS}[A]}},
\ee
where $S_{CS}[A]=\frac{1}{2}\int_0^1 d\tau\,\tr A_\tau$.

 \subsection{Gauge-fixing of the Yang-Mills field}\label{sec:gaugeYM}
 
For the Yang-Mills fields, the equivalent to the Lorenz gauge on the defect would be a condition on the pullback such that
\be\label{eq:lorenzA}
\partial_\tau A_\tau=0.
\ee
With this gauge fixing, the holonomy along $\cC$ is a constant matrix
\be
\oint_\cC dx^\mu A_\mu=\int_0^1d\tau\, A_\tau\equiv  \bar{A},
\ee
and the expectation value of the Wilson loop will depend only on the eigenvalues $a_i$ of the constant holonomy
\be\label{eq:expwconsta}
\vev{\cW(\cC)}=\vev{\sum_{i=1}^N e^{i a_i}}.
\ee
However, the Yang-Mills field lives on the whole spacetime, and not just at the defect. In order to fix the gauge appropriately, one should introduce a condition that is well-defined everywhere and that particularizes to \eqref{eq:lorenzA} when evaluated along the curve where the defect is extended. 

The simplest example where the gauge-fixing can be done is for a curve $\cC$ that is a circle of radius $R$ on a fixed plane in space. Without loss of generality, one can choose a coordinate system such that
\be
x(\tau)=\left(\begin{array}{cccc} 0, & R\cos(2\pi\tau), & R\sin(2\pi\tau), & 0\end{array}\right). 
\ee
The circle span by $\tau$ can be though of as part of an auxiliary four-dimensional space, parametrized by worldvolume coordinates $\sigma^\mu$ in such a way that $\tau$ corresponds to an angular direction, for instance the azimuthal angle of spatial spherical coordinates 
\be\label{eq:sigmaspher}
\sigma^0=t,\ \ \sigma^1=r\sin\theta\cos(2\pi\tau),\ \ \sigma^2=r\sin\theta\sin(2\pi\tau),\ \ \sigma^3=r\cos\theta.
\ee
The auxiliary space is mapped to the real space through a set of embedding functions $X^\mu(\sigma)$. For the simple case of the circular Wilson loop, these are
\be
X^\mu(\sigma)=\sigma^\mu.
\ee
In this case $\cC$ corresponds to the curve at $t=0$, $r=R$, $\theta=\pi/2$. The gauge-fixing condition for the Yang-Mills fields is defined over the whole spacetime using the embedding functions 
\be\label{eq:gaugecond}
0=\partial_\tau X^\mu \partial_\tau X^\nu \partial_\mu A_\nu[X]+\partial_\tau^2 X^\mu A_\mu[X]. 
\ee
Evaluating the gauge condition on $\cC$, one obtains the Lorenz gauge condition on the pullback
\be
0=\dot{x}^\mu \dot{x}^\nu \partial_\mu A_\nu[x]+\ddot{x}^\mu A_\mu[x]=\partial_\tau A_\tau(\tau).
\ee
If $x(\tau)$ is a more complicated curve, one should find first a smooth change of coordinates that maps it to a circle on a plane.
 
The gauge-fixing is introduced in the path integral following the BRST procedure
\be
\vev{\cW(\cC)}=\cN \int_0^{2\pi} da_0\, \int \cD \chi\cD \chi^\dagger\cD B^a\cD c^a\cD \bar{c}^a \;\vev{e^{i S_W+iS_{CS}+iS_{BRST}}},
\ee
where $B^a$ is the Nakanishi-Lautrup field and $c^a$, $\bar{c}^a$ are the ghost and anti-ghost fields, all in the adjoint representation of the group. The BRST-exact gauge-fixing action is
\be
S_{BRST}=\int d^4x\, \frac{1}{g^2}\left( B^a \left( \zeta^{\mu\nu}\partial_\mu  A_\nu +\eta^\mu A_\mu\right) -i\bar{c}^a \,\left( \zeta^{\mu\nu}\partial_\mu D_\nu c^a+\eta ^\mu D_\mu c^a\right)\right).
\ee
Where the tensors $\zeta^{\mu\nu}$ and $\eta^\mu$ are defined as
\be
\begin{split}
&\zeta^{\mu\nu}(x)=\int d^4\sigma\, \delta^{(4)}(x-X(\sigma))\,  \partial_\tau X^\mu \partial_\tau X^\nu,\\
&\eta^\mu(x)=\int d^4\sigma\, \delta^{(4)}(x-X(\sigma)) \, \partial_\tau^2 X^\mu.
\end{split}
\ee
Integrating out $B^a$ will fix the gauge to \eqref{eq:gaugecond}, but clearly the ghost action is highly non-trivial, as it is coupled to the Yang-Mills field and depends on the curve in this gauge. The integration over the ghost fields will be left implicit in the following. The defect action can be further simplified by doing a global $SU(N)$ transformation on the fermions $\chi\to U\chi$, such that the holonomy of the Yang-Mills field is diagonalized
\be
U^\dagger \bar{A} U=\bar{A}_D={\rm diag}(a_1,\cdots,a_N).
\ee
Then,  the completely gauge-fixed expectation value of the Wilson loop is
\be
\vev{\cW(\cC)}=\cN \vev{ \int_0^{2\pi} da_0\, \int \cD \chi\cD \chi^\dagger \;e^{i S_W[a_0,\bar{A}_D]+i k a_0+\frac{i}{2} \sum_{i=1}^N a_i}}_{g.f}.
\ee

\subsection{Expectation value of the Wilson loop}\label{sec:expvalw}

In the Lorenz gauge the action of the defect fermions only depends on the constant holonomies along the curve where the Wilson loop is defined. Since the action for the fermions is quadratic, integrating them out just introduces a determinant factor in the path integral
\be
\vev{\cW(\cC)}=\cN \vev{ \int_0^{2\pi} da_0\, \det((i\partial_\tau+a_0)\mathbb{1}+\bar{A}_D) e^{i k a_0+\frac{i}{2} \sum_{i=1}^N a_i} }_{g.f}.
\ee
The determinant can be evaluated using standard methods, the details can be found in Appendix~\ref{app:detferm}. The result is, up to a normalization factor that does not depend on the holonomies
\be\label{eq:measure}
\det((i\partial_\tau+a_0)\mathbb{1}+\bar{A}_D) \propto  e^{-i\frac{N}{2} a_0-\frac{i}{2} \sum_{j=1}^N a_j}\prod_{i=1}^N\left(1+e^{ ia_0+ i a_i} \right).
\ee
The overall phase factor corresponds to an anomalous contribution that cancels with the Chern-Simons term in the action, as advertised. Expanding the product and integrating over the holonomy of the defect gauge field, the result is
\be
\vev{\cW(\cC)}=\cN \vev{ \int_0^{2\pi} da_0  e^{- i a_0}\prod_{i=1}^N\left(1+e^{ ia_0+i a_i} \right)}_{g.f.} =\cN\vev{ \sum_{i=1}^N e^{ ia_i}}_{g.f.}.
\ee
Therefore, up to the undetermined constant normalization, the path integral over the defect fields indeed produces the expectation value for the Wilson loop \eqref{eq:expwconsta}.

\section{Weak coupling expansion of the defect theory}\label{sec:weak}

The calculation of the Wilson loop can be done by integrating out the Yang-Mills fields and working with the effective action of the fields at the defect. The defect action will be split in two parts, one corresponding to the interaction with the Yang-Mills fields and another involving only the defect fields
\be
\vev{\cW(\cC)}=\cN \int \cD \chi \cD\chi^\dagger \cD a_\tau \vev{e^{i \int J\cdot A}} e^{i\widetilde{S}_W+i\widetilde{S}_{CS}},
\ee
where the defect action in the path integral is
\be
\widetilde{S}_W=\int_0^1 d\tau\,\chi^\dagger\,i(\partial_\tau-i a_\tau)\chi, \ \ \widetilde{S}_{CS}=\int_0^1 d\tau\, k a_\tau,
\ee
and the interaction term is
\be
\int J\cdot A=\int d^4 x J^{a\mu} A_\mu^a,\ \ J^{a\mu}(x)=\int_0^1 d\tau\dot{x}^\mu\, j^a(\tau) \, \delta^{(4)}(x-x(\tau)).
\ee
The $U(N)$ worldline current $j^a$ is defined as
\be\label{eq:suncurr}
j^a=\chi^\dagger T^a \chi+\frac{\sqrt{N}}{2\sqrt{2}}\delta^{a0}.
\ee
The constant piece introduces in the defect action the Chern-Simons term for the Abelian component of the $U(N)$ gauge field. If the group is $SU(N)$, the constant piece is absent.

\begin{figure}
\hspace{-0.5cm}
\begin{tabular}{llllllll}

\begin{minipage}{2.5cm}
\includegraphics[width=2.5cm]{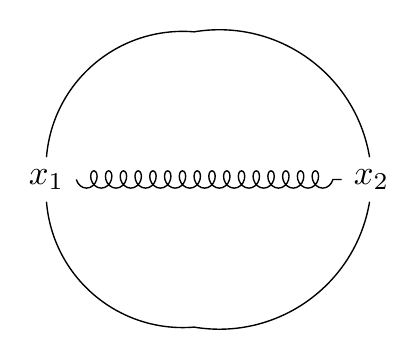}
\end{minipage}

&

\begin{minipage}{0.15cm}
{\Huge $+$}
\end{minipage}

&

\begin{minipage}{2.5cm}
\includegraphics[width=2.5cm]{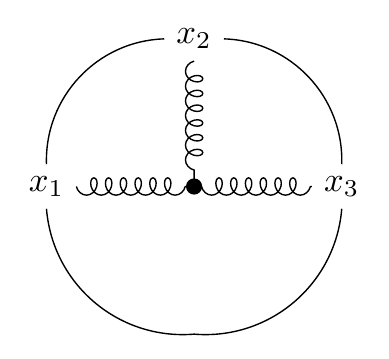}
\end{minipage}

&

\begin{minipage}{0.15cm}
{\Huge $+$}
\end{minipage}

&

\begin{minipage}{2.5cm}
\includegraphics[width=2.5cm]{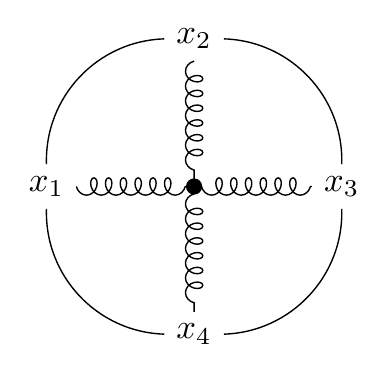}
\end{minipage}

&

\begin{minipage}{0.15cm}
{\Huge $+$}
\end{minipage}

&

\begin{minipage}{2.5cm}
\includegraphics[width=2.5cm]{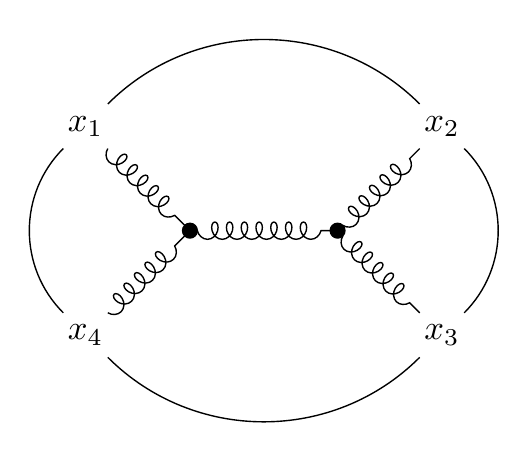}
\end{minipage}
&

\begin{minipage}{0.25cm}
{\Huge $+\cdots$}
\end{minipage}
\end{tabular}

\caption{\small Diagrammatic expansion of the action at the defect, up to the four-current vertex. The solid line represents the curve where the Wilson loop is supported. Each curly line represents a full gluon propagator and each thick dot a full vertex from the 1PI effective action.}\label{fig:weakexp}

\end{figure}

Integrating out the Yang-Mills fields will give the generating functional for an external current in the adjoint representation
\be
\vev{\cW(\cC)}=\cN \int \cD \chi \cD\chi^\dagger \cD a_\tau \,e^{i\widetilde{S}_W+i\widetilde{S}_{CS}+i W[J]}.
\ee
A proper calculation of the generating functional needs to take into account renormalization of the Yang-Mills theory. The renormalization scale will be fixed to a value $\mu$ such that the renormalized coupling $g$ is small enough to do a weak coupling expansion. In perturbation theory, $n$-point correlators of Yang-Mills fields start at least at $O(g^n)$, so the generating functional  admits a weak coupling expansion in terms of the connected time-ordered correlators of the Yang-Mills fields in vacuum 
\be
iW[J]=iW_0+i\int d^4 x\, \vev{A_\mu^a(x)} J^{a\mu}(x)+\frac{i^2}{2} \int\int d^4 x d^4y \, \vev{T(A_\mu^a(x)A_\nu^b(y))}_c J^{a\mu}(x) J^{b\nu}(y)+\cdots.
\ee
In perturbation theory Yang-Mills correlators can be computed systematically from tree-level diagrams involving the exact propagators and vertices obtained from the renormalized one-particle irreducible (1PI) action. The expansion only involves propagators and vertices of the gauge fields, as all the external legs attached to the Wilson loop in the diagrammatic expansion have to be gauge field propagators. The first terms in the expansion are drawn in Figure~\ref{fig:weakexp}.

Exact propagators and vertices are at least $O(g^{2(n-1)})$, but they also have an additional weak coupling expansion, so at each order in $g$ there can be several contributions from different connected correlators. Since  the constant term can be absorbed in the path integral normalization and $\vev{A_\mu^a(x)} =0$, the leading order correction is  $O(g^2)$.

To make formulas more compact a shorthand notation will be used for integrals and functions of the worldline coordinate
\be
\int_{12\cdots n}=\int_0^1 d\tau_1\int_0^1 d\tau_2\cdots\int_0^1 d\tau_n,\ \ f(\tau_1,\tau_2,\cdots,\tau_n)=f_{12\cdots n}.
\ee
From the point of view of the defect theory, the term corresponding to the $n$-point correlator in the expansion of the generating functional introduces a $n$-current vertex
\be\label{eq:W}
iW=\sum_{n=2}^\infty iW_n=\sum_{n=2}^\infty \frac{i^n}{n!}\int_{1\cdots n}K_{1\cdots n}^{(n)\,a_1\cdots a_n}  j_1^{a_1}\cdots j_n^{a_n} .
\ee
The kernels that determine the vertex between the currents in \eqref{eq:W} are 
\be\label{eq:kernels}
K^{(n)\,a_1\cdots a_n}_{1\cdots n}=\dot{x}_1^{\mu_1} \cdots \dot{x}_n^{\mu_n}G_{\mu_1\cdots\mu_n}^{a_1\cdots a_n}(x_1,\cdots,x_n),
\ee
where $G$ are the renormalized time-ordered connected correlators
\be
G_{\mu_1\cdots\mu_n}^{a_1\cdots a_n}(x_1,\cdots,x_n)=\vev{T(A_{\mu_1}^{a_1}(x_1)\cdots A_{\mu_n}^{a_n}(x_n))}_c.
\ee
Once the gauge for the defect field is fixed to the Lorenz gauge, the expectation value of the Wilson loop becomes
\be
\vev{\cW(\cC)}=\cN \int_0^{2\pi} da_0 e^{ ik a_0}\, \int \cD \chi \cD\chi^\dagger e^{i \widehat{S}_W+iW[J]} \, 
\ee
where
\be\label{eq:hatSW}
\widehat{S}_W=\int_0^1 d\tau\,\chi^\dagger\,i(\partial_\tau-i a_0)\chi.
\ee

\subsection{UV divergences and regularization of the defect action at leading order}

Assuming that the renormalization of the gauge coupling has been properly taken into account, the only divergence a Wilson loop defined on a smooth curve can have is linear in the cutoff $\Lambda$ and proportional to the length $L$ of the loop \cite{Polyakov:1980ca,Gervais:1979fv,Dotsenko:1979wb}
\be
\vev{\cW(\cC)} \sim e^{-\Lambda L}\ \times\, {\rm finite \, factors}. 
\ee
As the divergent part appears as an overall factor, it can be removed by multiplicative renormalization of the Wilson loop operator. Once this has been done, the resulting value is a finite function of the gauge coupling. Both the direct calculation of the Wilson loop and the defect action involve the pullbacks on the curve of the gauge field correlators. The same renormalization properties are then expected of the defect action, once the linear divergence has been taken care of, the remaining action depending on the renormalized gauge coupling should be finite. 

Since the vertices obtained from the 1PI renormalized action are UV finite, UV singularities can only appear when the points connected by an exact propagator in one of the connected correlators become coincident. Taking as an example the three-point connected correlator (second term in Fig.~\ref{fig:weakexp}), the correlator is determined by a diagram with three propagators starting at points $x_1$, $x_2$, $x_3$ on the curve $\cC$ and joining at a vertex at an arbitrary position in space $x$. One should integrate over all positions, so the position of the vertex can become coincident with any of the points at the curve, in which case the propagator becomes singular. However, for generic points on the curve the connected correlator is not singular, otherwise it would be singular when evaluated at  {\em any} three arbitrary points in spacetime. The same argument applies to higher order connected correlators, so the only possible singular contributions are when two or more points on the defect become coincident and a vertex comes close to the coincident points, or for the two-point connected correlator, when the two points on the curve become coincident. 

At leading order in the weak coupling expansion, the UV divergence is the one associated to the Yang-Mills field two-point correlator in the two-current vertex
\be
iW_2=\frac{i^2}{2} \int_{12}  K_{12}^{(2)\,a_1 a_2} j_1^{a_1} j_2^{a_2},
\ee
and it appears when the two points on the curve become coincident $x_1=x_2$. In the absence of self-intersections of the curve, this happens at equal values of the worldline coordinate $\tau_1=\tau_2$. The UV divergence can be regulated using a cutoff $\Lambda$ and allowing for the addition of counterterms that will remove the divergence when $\Lambda\to \infty$. The regulated two-current vertex is
\be
iW_2^\Lambda=\frac{i^2}{2} \int_{12} \int_{12}  K_{12}^{(2)\,a_1 a_2} j_1^{a_1} j_2^{a_2}\Theta\left( |x_1-x_2|-\frac{1}{\Lambda}\right)+ iS_{ct}.
\ee
The kernel $K^{(2)}$ in the two-current vertex is given in Eq.~\eqref{eq:kernels}. 

A straightforward computation (see Appendix~\ref{app:counter}) shows that the divergence is linear in the cutoff $\Lambda$. The defect action can be renormalized by adding a local counterterm that exactly cancels it. The counterterm in $R_\xi$ gauge is 
 \be\label{eq:sct}
iS_{ct}=\Lambda(\xi-3)\frac{(i g)^2}{4\pi^2} \int_0^1 d\tau\,  e\,j^a(\tau) j^a(\tau).
\ee
The UV divergence is gauge-dependent and vanishes in Yennie gauge $\xi=3$. This gauge has been used in some QED studies for its infrared properties. The Ward identity of the electromagnetic current fixes the relation between the wavefunction renormalization of the current and the renormalization of the vertex. In a general covariant gauge each of them has spurious IR divergences that are gauge-dependent and cancel out in physical quantities. In Yennie gauge, these IR divergences are absent, thus the wavefunction renormalization is IR finite \cite{Fried:1958zz,Yennie:1961ad}. This would translate into an IR finite renormalization of an infinitely extended Wilson line, which acts as an off-shell current, and it turns out it is also related to the cancellation of the UV divergence in the Wilson loop. Similar types of relations between UV and IR divergences have been known for a long time in the context of perturbative QCD amplitudes \cite{Ivanov:1985bk,Ivanov:1985np,Korchemsky:1985xj}.

\subsection{UV and IR finiteness in Yennie gauge}

\begin{figure}
\begin{center}
\begin{tabular}{ccc}
\begin{minipage}{2.5cm}
\includegraphics[width=2.5cm]{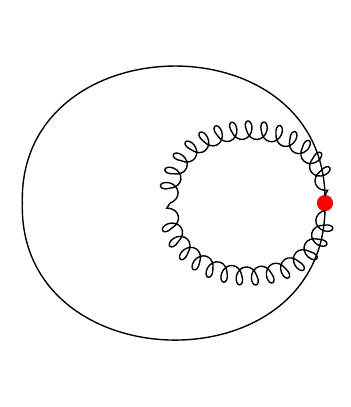}
\end{minipage}
&
{\Huge $\longrightarrow$}
&
\begin{minipage}{2cm}
\includegraphics[width=2.5cm]{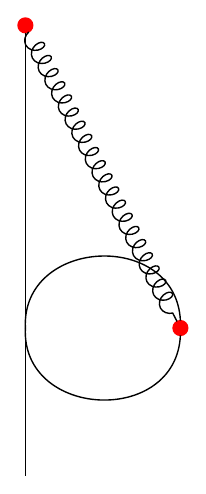}
\end{minipage}
\end{tabular}

\caption{An inversion maps a UV divergent diagram where the two endpoins of the correlator are very close on the Wilson loop curve to an IR divergent diagram where one of the endpoints is taken to infinity on the image of the Wilson loop curve.}\label{fig:inversion}
\end{center}
\end{figure}

The connection of the IR finiteness of Yennie gauge to the UV divergence in the Wilson loop defect action may be traced to the properties of the two-point correlator under inversion at the coincidence point
\be
(x_1-x_2)^\mu \to  \frac{(x_1-x_2)^\mu}{(x_1-x_2)^2}.
\ee
The defect action is schematically of the form  (color indices will be omitted in this discussion)
\be
\sim \int d\tau_1\,\dot{x}_1^\mu \int d\tau_2 \, \dot{x}_2^\nu G_{\mu\nu}(x_1-x_2)\, j_1 j_2.
\ee
For each fixed value of $x_1$ inside the $\tau_2$ integral, the coincident point can be taken to infinity by doing an inversion of the $x_2$ coordinate centered on $x_1$. The UV divergent diagram where two endpoints of the Yang-Mills propagator become coincident becomes an IR divergent diagram where one of the endpoints is at a point in the original curve and the other endpoint is taken to infinity along the trajectory resulting from the inversion (see Fig.~\ref{fig:inversion}). The IR properties of the propagator in the Yennie gauge are thus connected through the inversion to the absence of the UV divergence at coincident points.

Under the inversion, the defect action is transformed to
\be
\sim \int d\tau_1\,\dot{x}_1^\mu \int d\tau_2  \,\frac{\dot{x}_2^\alpha}{(x_1-x_2)^2}\,\cI_\alpha^{\ \nu}(x_1-x_2) G_{\mu\nu}\left(\frac{x_1-x_2}{(x_1-x_2)^2}\right)\, j_1 j_2.
\ee
where 
\be
\cI_{\mu\nu}(x_1-x_2)=\eta_{\mu\nu}-2\frac{(x_1-x_2)_\mu(x_1-x_2)_\nu}{(x_1-x_2)^2}.
\ee
Effectively, this amounts to a transformation of the two-point correlator
\be
\tilde{G}_{\mu\nu}(x_1-x_2)=(x_1-x_2)^2 \eta_{\mu\rho}\cI_{\nu\sigma}(x_1-x_2) G^{\rho\sigma}(x_1-x_2).
\ee
In Yennie gauge the two-point correlator is
\be
G_{\mu\nu}^Y(x_1-x_2)=\frac{1}{2\pi^2(x_1-x_2)^2}\left[\eta_{\mu\nu}-\frac{(x_1-x_2)_\mu(x_1-x_2)_\nu}{(x_1-x_2)^2}\right],
\ee
and satisfies the property of being transverse to the separation vector
\be
(x_1-x_2)^\nu G_{\mu\nu}^Y(x_1-x_2)=0.
\ee
Thanks to this, it transforms trivially under the inversion
\be
\tilde{G}_{\mu\nu}^Y(x_1-x_2)=(x_1-x_2)^2 G_{\mu\nu}^Y(x_1-x_2).
\ee
In other gauges, the two-point correlator is
\be
G_{\mu\nu}^\xi(x_1-x_2)=G_{\mu\nu}^Y(x_1-x_2) +\frac{\xi-3}{8\pi^2}\frac{\cI_{\mu\nu}(x_1-x_2)}{(x_1-x_2)^2}.
\ee
The tensor structure of the last term is not invariant under the inversion
\be
\tilde{G}_{\mu\nu}^\xi (x-y)=(x_1-x_2)^2 G_{\mu\nu}^Y(x_1-x_2)+\frac{\xi-3}{8\pi^2}\eta_{\mu\nu}.
\ee
The non-invariant term can be cast as a total derivative contribution using
\be
\partial_\mu^1\partial_\mu^2\,(x_1-x_2)^2 =-2\eta_{\mu\nu}.
\ee
Then, the transformed correlator is
\be\label{eq:transcorr}
\tilde{G}_{\mu\nu}^\xi (x-y)=(x_1-x_2)^2 G_{\mu\nu}^Y(x_1-x_2)-\frac{\xi-3}{16\pi^2}\partial_\mu^1 \partial_\mu^2\, (x_1-x_2)^2 .
\ee
In the limit $|x_1-x_2|\to \infty$, the contribution from the Yennie part of the transformed correlator is finite. The argument is simpler if one uses a coordinate system such that  the coincidence point is at the origin before the inversion, and it is approached along the trajectories
\be
x_1^\mu=v^\mu\tau+\frac{1}{2}a^\mu \tau^2+\cdots, \ \ x_2^\mu=-v^\mu\tau+\frac{1}{2}a^\mu \tau^2+\cdots.
\ee
The inversion changes $x_2^\mu\to x_2^\mu/(x_2)^2$,
\be
x_2^\mu=-\frac{v^\mu}{v^2}\frac{1}{\tau}-\frac{v^\mu}{v^2} \frac{(v\cdot a)}{v^2}+\frac{a^\mu}{2 v^2}.
\ee
An explicit calculation shows that $1/\tau^2$ and $1/\tau$ divergences cancel out in the contribution from the Yennie correlator
\be
K^{(2)}_{Y\,12}=\dot{x}_1^\mu \dot{x}_2^\mu (x_1-x_2)^2 G_{\mu\nu}^Y(x_1-x_2)\sim O(\tau^0).
\ee
For $\xi\neq 3$, the total derivative term in the transformed correlator \eqref{eq:transcorr} gives IR-divergent contributions localized at the endpoints introduced by the regulator. 

A few simple examples can illustrate the absence of UV divergences in Yennie gauge. Consider the trajectories
\be
\begin{array}{l|l}
\text{spatial line} &\ x^\mu=(0,0,0,L\tau),\\
\text{spatial circle} &\ x^\mu=(0,R\cos(2\pi\tau),R\sin(2\pi\tau),0),\\
\text{boosted} &\ x^\mu=\gamma (\tau,\beta\tau,0,0),\\
\text{accelerated} &\ x^\mu=(a\sinh\tau,a\cosh\tau,0,0).
\end{array}
\ee
The two-current kernel as defined in \eqref{eq:kernels} is, in each case,
\be\label{eq:kernelY}
\begin{array}{l|l}
\text{spatial line} & \ K^{(2)}_{12}=\frac{3-\xi}{8\pi^2 (\tau_1-\tau_2)^2},\\
\text{spatial circle} &\ K^{(2)}_{12}=-\frac{1}{2}+\frac{3-\xi}{8\sin^2\left(\pi(\tau_1-\tau_2) \right)},\\
\text{boosted} &\ K^{(2)}_{12}=\frac{3-\xi}{8\pi^2 (\tau_1-\tau_2)^2},\\
\text{accelerated} &\ K^{(2)}_{12}=\frac{1}{8\pi^2}+\frac{\xi-3}{32\pi^2\sinh^2\left(\frac{\tau_1-\tau_2}{2} \right)}.
\end{array}
\ee
All are finite in Yennie gauge $\xi=3$, and in fact the kernel for the spatial line and boosted trajectory vanishes, while for the spatial circle and the accelerated trajectory is a constant of opposite sign (the different magnitude just coming from the normalization of the worldline coordinate). Note however that only for the spatial circle the trajectory follows a closed curve of finite length, so the calculation of the Wilson loop using the defect action does not apply directly to the other examples. It would be interesting to study whether the analysis can be extended to those cases. 

The conclusion from the above discussion is that the UV divergence that appears when the endpoints of the propagator coincide is directly related through an inversion to spurious IR divergences like the ones observed in the wavefunction renormalization of the current. Working in the Yennie gauge the two-current vertex is manifestly finite and no regulator is needed, although in principle one could also use a different gauge and introduce the counterterm \eqref{eq:sct}. The finiteness properties of the propagator also suggests that UV singularities appearing at coincident points of $n$-point correlators in higher order terms might be dealt with in the same way, so the resulting effective action would finite in Yennie gauge.

\subsection{Subleading corrections}\label{sec:subl}

There are two possible corrections to the defect action at $O(g^4)$. The first is coming from the leading order contribution of the three-point correlator of the gauge fields, entering in the kernel of the three-current interaction (second term in Fig.~\ref{fig:weakexp}). The second is coming from the renormalization of the two-point function of the gauge field, and modifies the two-current interaction. At $O(g^6)$ there are contributions from both the renormalization of the two-point and three-point correlators, and the leading contributions of the three and four-point gauge field vertices to the four-point connected correlator (last two terms in Fig.~\ref{fig:weakexp}). The expansion goes on with further renormalization factors and new effective vertices involving more than four gauge fields at higher order. 

In each diagram, the bosonic nature of the gauge bosons should be manifest in the form of a symmetry of the connected correlator under the exchange of two endpoints at the Wilson loop. The color structure of diagrams with vertices is antisymmetric due to the properties of the structure constants. Then, the spacetime structure should be antisymmetric as well to make the total diagram symmetric. When all the points are coincident, the connected correlator is contracted with a tensor proportional to the product of velocities $\dot{x}^\mu$ at the coincident point. Since this tensor is symmetric and the connected correlator antsymmetric, the resulting contribution to the kernel vanishes. Therefore, the leading divergence when all the points of the diagram are coincident always vanishes. 

In the following it will be shown that all the potentially singular contributions to the three-current kernel vanish, so the resulting defect action is finite at $O(g^4)$. This gives some evidence in favor of the absolute finiteness of the defect action, but a complete systematic analysis of UV divergences will be deferred for future work. 

The connected three-point correlator is
\be
G_{\mu\nu\rho}^{abc}(x_1,x_2,x_3)=\frac{g^4}{8\pi^6} f^{abc}\int d^4 y \, \Gamma_{\mu\nu\rho}(x_1,x_2,x_3,y),
\ee
where
\be
\Gamma_{\mu\nu\rho}(x_1,x_2,x_3,y) =\sum_{\sigma \in S_3}\, \sign(\sigma) \Gamma^{\sigma(1)\sigma(2)\sigma(3)}_{\mu\nu\rho}(x_1,x_2,x_3,y).
\ee
The sum is over all possible permutations of pairs $(x_1,\mu)$, $(x_2,\nu)$, $(x_3,\rho)$, weighted by their sign, of the basic building block
\be
\Gamma^{123}_{\mu\nu\rho}(x_1,x_2,x_3,y)=\eta^{\alpha\beta}\partial_y^\gamma G_{\mu\alpha}(x_1-y) G_{\nu\beta}(x_2-y) G_{\rho\gamma}(x_3-y).
\ee
In this expression the gauge field propagators stripped of color and constant factors are, in Yennie gauge,
\be
G_{\mu\nu}(x-y)=  f(|x-y|) \left[ \eta_{\mu\nu}-\frac{(x-y)_\mu (x-y)_\nu}{(x-y)^2}\right],
\ee
where $f(|x-y|)=1/(x-y)^2$ to leading order in perturbation theory but can be a more general function if renormalization factors are taken into account. The following notation will be used for the separation vectors and the transverse projector
\be
u_i=x_i-y,\ \ P_{i\,\mu\nu}=\eta_{\mu\nu}-\frac{u_{i\,\mu} u_{i\,\nu}}{u_i^2}.
\ee
The derivative of the projector is
\be
\partial_y^\rho P_{i\,\mu\nu}=\frac{u_{i\,\nu} }{u_i^2}P_{i\, \mu}^{\ \ \rho}+\frac{u_{i\,\mu} }{u_i^2}P_{i\, \nu}^{\ \ \rho}.
\ee
Then, one finds the following expression for $\Gamma^{123}$ ($f_i=f(|u_i|)$)
\be\label{eq:g123}
\Gamma^{123}_{\mu\nu\rho}=f_1 f_2 f_3 \left[ \frac{1}{u_1^2}u_{1\,\mu} P_{2\,\nu}\cdot P_1\cdot P_{3\,\rho}+ \frac{f_1'}{|u_1|f_1} (P_{1\,\mu}\cdot P_{2\, \nu}) (u_1\cdot P_{3\,\rho})\right].
\ee
If two points, say $x_1$ and $x_2$, approach each other in a symmetric way
\be
x_1^\mu=x^\mu+v^\mu \sigma, \ \ x_2^\mu=x^\mu-v^\mu \sigma,\ \ x_3\neq x_1,x_2,
\ee
then, from \eqref{eq:g123}, the superficial degree of divergence at the two coincident points is logarithmic
\be
\int_{\sigma\to 0} d\sigma \int d^4 y \frac{1}{|v \sigma-y|^3 (v\sigma+y)^2}\underset{y= \sigma z}{\sim} \int d\sigma \frac{1}{\sigma} \int d^4 z \frac{1}{|v-z|^3(v+z)^2}\sim \log \sigma
\ee
However, the first term in \eqref{eq:g123} is symmetric under an odd $(23)$ permutation, and actually cancels out when the sum over all permutations is done for any three points $x_1$, $x_2$, $x_3$, as well as all other terms with similar structure. The second term  in \eqref{eq:g123} becomes symmetric under the odd $(12)$ permutation when $x_1=x_2$ and thus cancels out when the sum over all permutations is done. There is also a cancellation from the contraction of separation vectors with the projectors. Two terms survive in the sum
\be
\Gamma_{\mu\nu\rho}(x_1,x_1,x_3,y)= f_1 f_2 \frac{f_3'}{|u_3|}\left(  (P_{1\,\mu}\cdot P_{3\, \rho}) (u_3\cdot P_{1\,\nu})- (P_{1\,\nu}\cdot P_{3\, \rho}) (u_3\cdot P_{1\,\mu})\right).
\ee
This is an antisymmetric tensor in the $\mu\nu$ indices, so when contracted with the worldsheet velocities at the coincident points to compute the kernel, the result is vanishing
\be
\dot{x}_1^\mu\dot{x}_1^\nu \dot{x}_3^\rho\Gamma_{\mu\nu\rho}(x_1,x_1,x_3,y)=0.
\ee 
Therefore, the logarithmic divergence in the three-current kernel cancels out and the action is finite to $O(g^4)$. 

\section{Small circular Wilson loop}\label{sec:circleW}

Consider a spatial  Wilson loop defined on a circle of radius $R$
\be
x^\mu=(0,R\cos(2\pi\tau),R\sin(2\pi\tau),0).
\ee
In this case the einbein is constant and equal to the length of the circle $e=2\pi R$. If $R$ is small enough, the renormalization scale can be set to $\mu=1/R$, so theory remains weakly coupled $g\ll 1$ and perturbation theory is well behaved. The kernel of the two-current vertex in Yennie gauge is, from \eqref{eq:kernelY}
\be
K_{12}^{(2)\,a_1a_2}=-\frac{1}{2}\delta^{a_1a_2}.
\ee
The two-current vertex can be factored in the square of the integral of the current
\be\label{eq:W2circle}
iW_2=-\frac{(ig)^2}{4} \int_{12} j_1^a j_2^a=\frac{g^2}{4} \left(\int_0^1 d\tau j^a(\tau) \right)^2.
\ee
The Wilson loop to leading order in the weak coupling expansion is 
\be
\vev{\cW(\cC)}=\cN\int_0^{2\pi} da_0 e^{i k a_0} \int \cD\chi\cD\chi^\dagger e^{i\widehat{S}_W+i W_2 },
\ee
where $\widehat{S}_W$ was given in Eq.~\eqref{eq:hatSW}. Up to a constant normalization, the two-current vertex can be manipulated to an integral over hermitian matrices. This is done in Appendix~\ref{app:matrix}, with the result
\be
e^{i W_2}\propto  \int[dM] e^{iS_M}\propto \int \left(\prod_{i=1}^N dM_i\right)  \Delta^2(M) \, \delta_{(\mathfrak{s}) \mathfrak{u}}(M)\, e^{iS_M},
\ee
where $\Delta^2(M)$ factor is the Vandermonde determinant 
\be
\Delta^2(M)=\prod_{i<j} (M_i-M_j)^2,
\ee
and the factor $ \delta_{(\mathfrak{s})\mathfrak{u}}(M)$ imposes the tracelessness condition when the group is $SU(N)$.
\be
 \delta_{\mathfrak{u}}(M)=1,\ \ \delta_\mathfrak{su}(M)=\delta(\tr M)=\delta\left(\sum_{j=1}^N M_j\right).
\ee
The action in the matrix integral depends only on the eigenvalues after a unitary rotation of the fermions ($M_D={\rm diag}(M_1,\cdots,M_N)$)
\be\label{eq:slamb}
iS_M=-g\int d\tau\,\chi^\dagger M_D \chi -2\tr M_D^2-\frac{g}{2}\tr M_D.
\ee
With these expressions, the fermion action is quadratic and they can be integrated out to give a determinant term of the same form as \eqref{eq:measure} 
\be
\vev{\cW(\cC)}=\cN\int_0^{2\pi} da_0 e^{i k a_0}   \int [dM]  e^{-2\tr M_D^2-\frac{g}{2}\tr M_D}  \det\left((i\partial_\tau+a_0)\mathbb{1}+ig\bar{M}_D\right).
\ee
Following the same steps as in section \ref{sec:expvalw}, the integral over $a_0$ will give 
\be\label{eq:wilsonloopmat}
\vev{\cW(\cC)}=\cN\int \left(\prod_{i=1}^N dM_i\right)\Delta^2(M)\delta_{(\mathfrak{s})\mathfrak{u}}(M)\left(  \sum_{i=1}^N\,e^{-g M_i} \right)\,\  \,e^{-2\sum_{j=1}^N M_j^2}.
\ee
The anomalous contribution from the fermions has been cancelled with the term proportional to $\tr M_D$ in \eqref{eq:slamb}. Remarkably, the same expression was found for the supersymmetric Wilson loop \cite{Erickson:2000af,Drukker:2000rr,Pestun:2007rz}, but in that case it is exact, while here it is only valid to leading order in the weak coupling expansion. The value of the $U(N)$ Wilson loop can then be read directly from the supersymmetric result. 

Normalizing by the zero coupling value one finds for the $U(N)$ Wilson loop \cite{Drukker:2000rr} ($\lambda=g^2 N$ is the 't Hooft coupling)
\be\label{eq:circwil}
\mathfrak{w}_{U(N)}(\lambda)\equiv\frac{\vev{\cW(\cC)}_\lambda}{\vev{\cW(\cC)}_0}=\frac{1}{N} L_{N-1}^1\left(-\frac{\lambda}{4N} \right)e^{\frac{\lambda}{8N}},
\ee
where $L_n^m(x)$ is a generalized Laguerre polynomial. 

The $SU(N)$ matrix integral can be done by using the representation of the delta function\footnote{Another way to do it is separating the Abelian and non-Abelian parts before going to the eigenvalue variables, see e.g.\cite{Bourget:2018obm}.}
\be
\delta_{\mathfrak{su}}(M)=\frac{1}{2\pi}\int_{-\infty}^\infty du \, e^{iu\sum_{i=1}^N M_i}.
\ee
Completing squares and shifting all eigenvalues by the same constant $M_i\to M_i+i u/4$ gives
\be
\vev{\cW(\cC)}_{SU(N)}\propto \left(\int_{-\infty}^\infty du\, e^{-N u^2/8-i g u/4}\right)\vev{\cW(\cC)}_{U(N)}\propto e^{-g^2/(8N)}\vev{\cW(\cC)}_{U(N)}.
\ee
This gives the general formula
\be
\mathfrak{w}_{SU(N)}(\lambda)=e^{-\lambda/(8N^2)}\,\mathfrak{w}_{U(N)}(\lambda).
\ee
The first orders in the weak coupling expansion are
\be
\mathfrak{w}_{SU(N)}(\lambda)=1+\left(\frac{N^2-1}{2N}\right) \frac{\lambda}{4N}+\left(\frac{N^2-1}{2N}\right) \left(\frac{2N^2-3}{12 N} \right)\frac{\lambda^2}{16 N^2}+\cdots.
\ee
For a fixed value of the 't Hooft coupling, the first terms in the large-$N$ expansion are
\be
\mathfrak{w}_{SU(N)}(\lambda)=\frac{2}{\sqrt{\lambda}}I_1(\sqrt{\lambda})+\frac{\sqrt{\lambda}}{4N^2}\left(\frac{\sqrt{\lambda}}{12}I_2(\sqrt{\lambda})-I_1(\sqrt{\lambda})\right)+\cdots,
\ee
where $I_n(x)$ are Bessel functions. At very large values of the coupling $\lambda\to \infty$, the leading behavior of the planar term is $I_1(\sqrt{\lambda})\sim e^{\sqrt{\lambda}}$, which is of the form found in AdS/CFT calculations \cite{Berenstein:1998ij,Drukker:1999zq}.

\subsection{Radial dependence}

The final result for the circular Wilson loop \eqref{eq:circwil} does not depend explicitly on the radius but, for a non-conformal theory, there is an implicit dependence through the running of the coupling constant. At one loop the beta function of the 't Hooft coupling is 
\be
\beta(\lambda)=-\beta_0 \frac{\lambda^2}{8\pi^2N},
\ee
where $\beta_0> 0$ is a scheme-independent constant coefficient that depends on the matter content of the theory. The running coupling evaluated at the scale $\mu=1/R$ is
\be\label{eq:grun}
\lambda=-\frac{16\pi^2 N}{\beta_0}\frac{1}{\log\left( (\Lambda_{YM} R)^2\right)}, \ \ \Lambda_{YM} R \ll 1.
\ee
The renormalization invariant scale $\Lambda_{YM}$ determines the size of the loop at which the weak coupling expansion breaks down. The Wilson loop is defined on a smooth curve, so it obeys the usual Callan-Symanzik equation: the change of the Wilson loop with the radius is determined by the beta function of the gauge coupling times an additional factor obtained from the derivative with respect to the coupling
\be
\begin{split}
-R\frac{\partial}{\partial R}\log\mathfrak{w}_{SU(N)}&=\frac{\beta(\lambda)}{8N^2}  \left[ N-1+2N\frac{L_{N-2}^2\left(-\frac{\lambda}{4N} \right)}{L_{N-1}^1\left(-\frac{\lambda}{4N}\right)}\right]
=\frac{\beta(\lambda)}{4N} \frac{N^2-1}{2N}F(\lambda).
\end{split}
\ee
The normalization $F(0)=1$  has been chosen to highlight that the Wilson loop beta function is proportional to the Casimir of the fundamental representation of $SU(N)$, $C_2(N)=(N^2-1)/(2N)$,  at leading order. $F(\lambda)$ is a rational function of the 't Hooft coupling and has the following expansion 
\be
F(\lambda)=1+\frac{1}{4}\left(\frac{1}{3}-\frac{1}{2N^2}\right)\lambda+\frac{1}{128}\left(\frac{1}{3}-\frac{N^2-1}{N^4}\right)\lambda^2+\cdots.
\ee

\section{Summary and discussion}\label{sec:conc}

One of the main findings of this work is that ordinary Wilson loops share some of the nice features of their supersymmetric cousins. If the loop is defined on a smooth closed curve and the right gauge is chosen, they are both free of divergences to leading order in the weak coupling expansion.  The expectation value of $(S)U(N)$ Wilson loops can be computed from an effective theory of fields localized on a defect along the loop. Most likely this can be extended to other groups and different representations, taking as guidance the D-brane actions that describe supersymmetric Wilson loops (e.g. in \cite{Gomis:2006sb}). The effective defect action can be constructed systematically from connected tree-level diagrams of the exact gauge field vertices and propagators obtained from the renormalized 1PI action. 

A Wilson loop should be finite except for a possible linear divergence \cite{Polyakov:1980ca,Gervais:1979fv,Dotsenko:1979wb}. The linear divergence is removed from the defect action in Yennie gauge, so the resulting action is finite at $O(g^2)$ and, provided  the renormalization properties of the Wilson loop hold for the defect action, it would be expected to be finite at all orders. Some partial evidence is that the color structure implies that the would-be most divergent terms should be vanishing, and an explicit check shows that the $O(g^4)$ defect action is indeed free of divergences.

Besides a generalization in terms of groups and representations, interesting extensions would be to construct a defect action for Wilson loops on curves that are not smooth everywhere and curves that are not bounded to a finite region. In both cases one can extract interesting physics. Curves with cusps have UV divergences that modify the Callan-Symanzik equation of the Wilson loop, so its evolution is not determined uniquely by the running of the coupling \cite{Korchemsky:1987wg}. The cusp anomalous dimension determines the behavior of IR divergences in scattering amplitudes \cite{Korchemsky:1991zp,Korchemskaya:1996je}. The relation extends even beyond in the case of $\cN=4$ SYM. Using the holographic dual, it was found that finite terms in the amplitudes can also be obtained from the expectation value of a Wilson loop \cite{Alday:2007hr}. Exact results for BPS loops \cite{Fiol:2011zg,Correa:2012at,Fiol:2012sg} show that the cusp anomalous dimension in $\cN=4$ SYM is also related to the radiation emitted by an accelerated charged particle as well.  Another interesting result for BPS loops in $\cN=4$ SYM is that the expectation value of the circular and Polyakov loops are proportional to each other \cite{Mueck:2010ja}. Exploring generalizations of these results to non-BPS loops would be a very interesting direction to follow.

Using the defect action, the expectation value of the Wilson loop on a small circle was fairly easy to compute at leading order. The only interaction at $O(g^2)$ is a quadratic term for two defect currents, that factorizes in the square of the integral of the current. This allows to convert the path integral over the defect into a matrix integral, and it is found that the expectation value is universal as a function of the coupling and the rank of the group and coincides with the value of the $1/2$ BPS loop in $\cN=4$ SYM. In general, subleading corrections will spoil the factorization of the currents. This can already be seen in the term with two currents, as the renormalization of the coupling will introduce a term depending on the separation of the currents along the loop. However, if the theory is conformal, there are no renormalization factors and the $O(g^2)$ term will keep the leading order structure. In any case, the effective theory is simple enough that it may be possible to use it to compute subleading corrections to the expectation value of the Wilson loop and even extend it to more complicated curves. Weak coupling corrections to the circular Wilson loop  in $\cN=4$ SYM have been computed in \cite{Beccaria:2017rbe,Beccaria:2018ocq}. The value is indeed the same as the BPS loop at $O(g^2)$ but it deviates at $O(g^4)$, as expected.

Even though the analysis has been restricted to weak coupling, it is tempting to try to extrapolate some of the results to strong coupling and make some speculations. The leading order result can be though of as a resummation of all Feynman diagrams that do not have internal vertices, i.e. involve only gauge fields propagating from one point to another on the Wilson loop, and the loop corrections that enter in the renormalization. The fact that the Wilson loop expectation value is the same as for the BPS operator implies that, in the large-$N$ and strong 't Hooft coupling limit, it reproduces the characteristic behavior of Wilson loops computed in AdS/CFT
\be
\vev{\cW(\cC)}\sim e^{\sqrt{\lambda}}.
\ee
This suggests that the there could be a string theory dual description of the Wilson loop in the ``non-interacting'' sector of all $(S)U(N)$ gauge theories. The dual might be not a full-fledged string theory, but some truncated version, possibly one in which string interactions have been turned off but worldsheets of different topology are included. The weak coupling calculation would then contain information about the free string theory dual in a highly curved space. If the duality really exists, then subleading corrections to the Wilson loop would take into account string interactions. 

Another interesting connection to holography is through the worldsheet geometry. The holographic dual to the circular BPS loop is a surface that covers an $AdS_2$ region of the full geometry. It has been proposed that the Wilson loop has a similar dual description with different boundary conditions for the fields living on the surface \cite{Alday:2007he}\footnote{In this case, the Wilson and BPS loops would be further related by an RG flow, such that the coefficient of the coupling to the scalar in \eqref{eq:bpsW} runs with the scale \cite{Polchinski:2011im}, see also  \cite{Beccaria:2017rbe,Beccaria:2018ocq} for more evidence at weak coupling.}. This points to a relation of the defect theory to the Sachdev-Ye-Kitaev (SYK) model of one-dimensional fermions \cite{Sachdev:1992fk,Kitaev:2015,Polchinski:2016xgd}. Although there are some differences with SYK models, among others the absence of disorder, the classical Wilson loop defect action is invariant under worldline reparametrizations. In the quantum theory the symmetry could be broken both explicitly\footnote{For instance, fixing Lorenz gauge for the Abelian gauge field at the defect.} and spontaneously. Then, the arguments that determine the low energy effective action of the SYK model \cite{Maldacena:2016hyu} would apply to the defect action as well. The effective action that would result from this breaking can be connected directly to two-dimensional dilaton gravity in $AdS_2$ \cite{Almheiri:2014cka,Jensen:2016pah,Maldacena:2016upp}. It should also be noted that there are similar defect theories that have been proposed as models of quantum impurities in strongly correlated systems with an $AdS_2$ dual, see e.g. the reviews \cite{Sachdev:2010uj,Erdmenger:2015xpq} and references therein.

\section*{Acknowledgments}

I want to thank Antoine Bourget and Diego Rodriguez-Gomez for useful discussions, and to Prem Kumar and Horatiu Nastase for helpful comments. This work has been partially supported by the Spanish grant MINECO-16-FPA2015-63667-P, the Ramon y Cajal fellowship RYC-2012-10370 and GRUPIN 14-108 research grant from Principado de Asturias.

\appendix

\section{Calculation of the fermion determinant}\label{app:detferm}

The defect fermions satisfy antiperiodic boundary conditions, $\chi(1)=-\chi(0)$, so the Fourier expansion relative to the worldline coordinate is
\be
\chi(\tau)=\sum_{n=-\infty}^\infty \chi_n e^{-2\pi i \left(n+\frac{1}{2}\right)\tau}.
\ee
The fermionic determinant has a formal expression as an infinite product over the Fourier modes and color. Zeta-function regularization is assumed, so the determinant can be manipulated to
\be
\begin{split}
&\det\left( i(\partial_\tau+a_0)\mathbb{1} +  \bar{A}_D\right)=\prod_{i=1}^N \prod_{n=-\infty}^\infty \left(2\pi\left(n+\frac{1}{2}\right)+a_0 +a_i\right)\\
&=\prod_{i=1}^N \left[(\pi+a_0+a_i)\prod_{n=1}^\infty \left(-(2\pi n)^2+\left(\pi+a_0 +a_i\right)^2\right)\right]\\
&=\prod_{i=1}^N \left[(\pi+a_0+a_i)\prod_{n=1}^\infty \left(1-\frac{\left(\pi+a_0 +a_i\right)^2}{(2\pi n)^2}\right) \prod_{m=1}^\infty (-(2\pi m)^2)\right].
\end{split}
\ee
All the factors are manifestly finite except the last infinite product, which is defined using the Riemmann $\zeta$ function
\be
\zeta(s)=\sum_{n=1}^\infty \frac{1}{n^s}, \ \ \zeta'(s)=-\sum_{n=1}^\infty \frac{1}{n^s}\log n.
\ee
Its regularized value is
\be
\prod_{m=1}^\infty (-(2\pi m)^2)=\exp\left( \sum_{m=1}^\infty \log (-(2\pi m)^2) \right)=\exp\left( -2 \zeta'(0)+\left(2\log(2\pi)+i\pi \right)\zeta(0)\right). 
\ee
As $\zeta(0)=-1/2$ and $\zeta'(0)=-1/2\log(2\pi)$, the infinite product reduces to phase factor $e^{-iN\pi/2}$ that will be absorbed in the normalization of the Wilson loop. The integration over defect ghosts produces a similar factor.

The value of the regularized determinant becomes
\be
\begin{split}
&\det\left( i(\partial_\tau+a_0)\mathbb{1} +  \bar{A}_D\right)=\prod_{i=1}^N \left[(\pi+a_0+a_i)\prod_{n=1}^\infty \left(1-\frac{\left(\pi+a_0 +a_i\right)^2}{(2\pi n)^2}\right)\right]\\
&=\prod_{i=1}^N 2\cos\left(\frac{a_0 +a_i}{2} \right)=e^{-i\frac{N}{2}a_0-\frac{i}{2}\sum_{j=1}^N a_j}\prod_{i=1}^N (1+e^{ia_0+ia_i}).
\end{split}
\ee

\section{Calculation of the divergence in the two-current vertex}\label{app:counter}

The connected two-point correlator is, in $R_\xi$ gauge and to leading order in the weak coupling expansion, 
\be
\begin{split}
&G^{a_1a_2}_{\mu_1\mu_2}(x_1,x_2)= \frac{g^2}{4\pi^2}\left[\frac{1}{(x_1-x_2)^2}\eta_{\mu_1\mu_2}+\frac{1-\xi}{2}\frac{\partial}{\partial x_1^{\mu_1}}\frac{\partial}{\partial x_2^{\mu_2}}\log(\mu|x_1-x_2|)\right]\delta^{a_1 a_2}\\
&= \frac{g^2}{8\pi^2}\left[\frac{1+\xi}{(x_1-x_2)^2}\eta_{\mu_1\mu_2}+2(1-\xi)\frac{(x_1-x_2)_{\mu_1}(x_1-x_2)_{\mu_2}}{((x_1-x_2)^2)^2}\right]\delta^{a_1 a_2}.
\end{split}
\ee
The two-point correlator diverges when $|x_1-x_2|\to 0$. One can separate the interval around the singular point introducing a second scale  $\mu<\Lambda$
\be
\Theta\left( |x_1-x_2|-\frac{1}{\Lambda}\right)=\Theta\left( |x_1-x_2|-\frac{1}{\mu}\right)+\Theta\left( |x_1-x_2|-\frac{1}{\Lambda}\right)\Theta\left(  \frac{1}{\mu}-|x_1-x_2|\right).
\ee
The two-current term is split in a finite $\mu$-dependent part and the $\Lambda$-dependent contribution
\be
iW_2^{\Lambda}=iW_2^{\mu}+i\widetilde{W}_2^{\Lambda}+iS_{ct},
\ee
where the divergent piece is
\be
i\widetilde{W}_2^{\Lambda}= \int_{12}  K_{12}^{(2)\,a_1 a_2} j_1^{a_1} j_2^{a_2}\Theta\left( |x_1-x_2|-\frac{1}{\Lambda}\right)\Theta\left(  \frac{1}{\mu}-|x_1-x_2|\right).
\ee
In order to evaluate the divergence of the kernel it will be convenient to introduce symmetric coordinates
\be\label{eq:symcoord}
\tau=\frac{\tau_1+\tau_2}{2}, \ \ \sigma=\tau_1-\tau_2.
\ee
The divergence happens at small values of $\sigma$, where one can use the following expansions
\be\label{eq:symexp}
\begin{split}
&x^\mu_1=x^\mu(\tau)+\frac{1}{2}\dot{x}^\mu(\tau)\sigma+O(\sigma^2), \ \ x^\mu_2=x^\mu(\tau)-\frac{1}{2}\dot{x}^\mu(\tau)\sigma+O(\sigma^2),\\
&|x_1-x_2|=e|\sigma|+O(\sigma^3),\ \ j_1^a j_2^a=j^a(\tau) j^a(\tau)+O(\sigma^2).
\end{split}
\ee
Here $e=|\dot{x}(\tau)|$ is the worldline einbein for the $\tau$ coordinate. The two possible singular contributions to the kernel are
\be
\frac{(\dot{x}_1\cdot \dot{x}_2)}{(x_1-x_2)^2}=\frac{1}{\sigma^2}+O(1), \ \ \frac{(\dot{x}_1\cdot(x_1-x_2))(\dot{x}_2\cdot(x_1-x_2)) }{((x_1-x_2)^2)^2}=\frac{1}{\sigma^2}+O(1).
\ee
Adding all together,
\be
K_{12}^{(2),a_1,a_2}=\frac{3-\xi}{\sigma^2}\delta^{a_1 a_2}+O(1).
\ee
Expanding for $e\mu \gg 1$, the leading term is a linear divergence 
\be
\begin{split}
&i\widetilde{W}_2^{\Lambda}= (3-\xi)\frac{(ig)^2}{8\pi^2}\int d\tau j^a(\tau)j^a(\tau) \int d\sigma \left(\frac{1}{\sigma^2}+O(1)\right)  \Theta\left( e|\sigma|-\frac{1}{\Lambda}\right)\Theta\left(  \frac{1}{\mu}-e|\sigma|\right)\\
&=(3-\xi)\frac{(ig)^2}{4\pi^2}(\Lambda-\mu)\int_0^1 d\tau\, e j^a(\tau)j^a(\tau)  +O\left(\frac{1}{\Lambda},\frac{1}{\mu} \right).
\end{split}
\ee

\section{Transformation of the two-current vertex to a matrix integral}\label{app:matrix}

The calculation will be done for an {\em imaginary} coupling $g=-iz$ and then analytic continuation will be used to obtain the result for real values. This will be justified by the final result, that is analytic on the whole complex plane. The quartic term \eqref{eq:W2circle} equals an integration over a set of constant $N\times N$ hermitian matrices $\Sigma$. 
If the group is $SU(N)$, then the integral is restricted to traceless matrices. For any hermitian matrix $M$ the following measure factor is introduced 
\be
\delta_{\mathfrak{u}}(M)=1,\ \ \delta_\mathfrak{su}(M)=\delta(\tr M).
\ee
The two-current factor is
\be
e^{i W_2}=\int [d\Sigma]  \delta_{(\mathfrak{s})\mathfrak{u}}(\Sigma)\delta[\Sigma-\cO_{(\mathfrak{s})\mathfrak{u}}] e^{-\frac{1}{4}\left( \tr(T^a\Sigma)\right)^2 },
\ee
where 
\be
\begin{split}
\delta[\Sigma-\cO_\mathfrak{u}] =&\prod_{i,j}\delta\left[\Sigma_{ij}-z\int d\tau\left(\chi^\dagger_i\chi_j+\frac{1}{2}\delta_{ij}\right) \right],\\
\delta[\Sigma-\cO_\mathfrak{su}] =&\prod_{i,j}\delta\left[\Sigma_{ij}-z\int d\tau\left(\chi^\dagger_i\chi_j-\frac{1}{N}\delta_{ij}(\chi^\dagger \chi)\right) \right].
\end{split}
\ee
One can check easily that integrating over $\Sigma$ with the delta function gives back the original path integral with the current squared term. The delta function has a path integral representation in terms of a hermitian matrix $M$
\be
\begin{split}
\delta[\Sigma-\cO_{\mathfrak{u}}] =&\int [dM]\delta_{\mathfrak{u}}(M)\exp\left[-iM_{ij}\left(\Sigma_{ij}-z\int d\tau\left(\chi^\dagger_i\chi_j+\frac{1}{2}\delta_{ij}\right) \right) \right],\\
\delta[\Sigma-\cO_\mathfrak{su}] =&\int [dM]\delta_{\mathfrak{su}}(M)\exp\left[-iM_{ij}\left(\Sigma_{ij}-z\int d\tau\left(\chi^\dagger_i\chi_j-\frac{1}{N}\delta_{ij}(\chi^\dagger\chi)\right) \right) \right].
\end{split}
\ee
The quartic term is thus (all the terms proportional to $\tr M$ vanish in the case of $SU(N)$ group)
\be
e^{i W_2}=\int [dM][d\Sigma] \, \delta_{(\mathfrak{s})\mathfrak{u}}(\Sigma)\delta_{\mathfrak{su}}(M)\,e^{iS_\Sigma+\frac{iz}{2}\tr M}.
\ee
In this case the action for $\Sigma$ is gaussian and the path integral can be done explicitly. Using the identity for $SU(N)$ generators
\be
\sum_{a=1}^{N^2-1} T^a_{ij}T^a_{kl}=\frac{1}{2}\left(\delta_{il}\delta_{kj}-\frac{1}{N}\delta_{ij}\delta_{kl}\right),
\ee
the quadratic term is
\be
\left( \tr(T^a\Sigma)\right)^2=\frac{1}{2}\tr \Sigma^2+\left(1-\frac{1}{2N}\right)(\tr \Sigma)^2,
\ee
During the calculation matrices $X$ will be split in traceless $X_t$ and trace $\tr X$ parts
\be
X=X_t+\frac{1}{N}\tr X,\ \ \tr X_t=0.
\ee
This is used to complete squares in the action of the matrix integral
\be
\begin{split}
&iS_\Sigma=-\frac{1}{4}\left(\frac{1}{2}\tr \Sigma^2+\left(1-\frac{1}{2N}\right)(\tr \Sigma)^2 \right)-i\tr(\Sigma M)=\\
&-\frac{1}{8}\tr \Sigma_t^2-i\tr(\Sigma_tM_t)-\frac{1}{8}(\tr \Sigma)^2-\frac{i}{N}\tr\Sigma\tr M\\
&=-\frac{1}{8}\tr \left(\Sigma_t+4iM_t\right)^2-\frac{1}{8}\left(\tr \Sigma+\frac{ 4i}{N}\tr M\right)^2-2\tr M_t^2-\frac{2}{N^2}(\tr M)^2 \\
&=-\frac{1}{8}\tr \left(\Sigma+4iM\right)^2-2\tr M^2
\end{split}
\ee
The integral over $\Sigma$ contributes with just an overall constant factor that will be absorbed in the normalization of the Wilson loop. Regarding the integral over $M$,  any hermitian matrix can be written as a unitary rotation of a diagonal matrix $M=U M_D U^\dagger$, $M_D={\rm diag}\,(M_i,\cdots,M_N)$. The integral over hermitian matrices can be split in the usual way in the integral over eigenvalues times the unitary transformations 
\be
\int [dM] =\int [dU] \int \prod_i dM_i \Delta^2(M),
\ee
where the Vandermonde determinant that appears in the measure of the eigenvalues is
\be
\Delta^2(M)=\prod_{i<j} (M_i-M_j)^2.
\ee
By doing a global $SU(N)$ rotation of the fermions, $\chi\to U\chi$, the action becomes independent of the unitary matrices $U$, whose integral will just give a constant factor proportional to the volume of the group. Then,
\be
e^{i W_2}\propto  \int \prod_i dM_i \Delta^2(M)\, \delta_{\mathfrak{su}}(M)\, e^{iS_M}.
\ee
Doing the analytic continuation to real values of the coupling,
\be
iS_M=-g\int d\tau\,\chi^\dagger M_D \chi -2\tr M_D^2-\frac{g}{2}\tr M_D.
\ee

\bibliographystyle{JHEP}
\bibliography{biblio}

\providecommand{\href}[2]{#2}\begingroup\raggedright\begin{thebibliography}{10}

\bibitem{Erickson:2000af}
J.~K. Erickson, G.~W. Semenoff and K.~Zarembo, \emph{{Wilson loops in N=4
  supersymmetric Yang-Mills theory}},
  \href{https://doi.org/10.1016/S0550-3213(00)00300-X}{\emph{Nucl. Phys.}
  {\bfseries B582} (2000) 155}
  [\href{https://arxiv.org/abs/hep-th/0003055}{{\ttfamily hep-th/0003055}}].

\bibitem{Drukker:2000rr}
N.~Drukker and D.~J. Gross, \emph{{An Exact prediction of N=4 SUSYM theory for
  string theory}}, \href{https://doi.org/10.1063/1.1372177}{\emph{J. Math.
  Phys.} {\bfseries 42} (2001) 2896}
  [\href{https://arxiv.org/abs/hep-th/0010274}{{\ttfamily hep-th/0010274}}].

\bibitem{Pestun:2007rz}
V.~Pestun, \emph{{Localization of gauge theory on a four-sphere and
  supersymmetric Wilson loops}},
  \href{https://doi.org/10.1007/s00220-012-1485-0}{\emph{Commun. Math. Phys.}
  {\bfseries 313} (2012) 71} [\href{https://arxiv.org/abs/0712.2824}{{\ttfamily
  0712.2824}}].

\bibitem{Berenstein:1998ij}
D.~E. Berenstein, R.~Corrado, W.~Fischler and J.~M. Maldacena, \emph{{The
  Operator product expansion for Wilson loops and surfaces in the large N
  limit}}, \href{https://doi.org/10.1103/PhysRevD.59.105023}{\emph{Phys. Rev.}
  {\bfseries D59} (1999) 105023}
  [\href{https://arxiv.org/abs/hep-th/9809188}{{\ttfamily hep-th/9809188}}].

\bibitem{Drukker:1999zq}
N.~Drukker, D.~J. Gross and H.~Ooguri, \emph{{Wilson loops and minimal
  surfaces}}, \href{https://doi.org/10.1103/PhysRevD.60.125006}{\emph{Phys.
  Rev.} {\bfseries D60} (1999) 125006}
  [\href{https://arxiv.org/abs/hep-th/9904191}{{\ttfamily hep-th/9904191}}].

\bibitem{Maldacena:1998im}
J.~M. Maldacena, \emph{{Wilson loops in large N field theories}},
  \href{https://doi.org/10.1103/PhysRevLett.80.4859}{\emph{Phys. Rev. Lett.}
  {\bfseries 80} (1998) 4859}
  [\href{https://arxiv.org/abs/hep-th/9803002}{{\ttfamily hep-th/9803002}}].

\bibitem{Rey:1998ik}
S.-J. Rey and J.-T. Yee, \emph{{Macroscopic strings as heavy quarks in large N
  gauge theory and anti-de Sitter supergravity}},
  \href{https://doi.org/10.1007/s100520100799}{\emph{Eur. Phys. J.} {\bfseries
  C22} (2001) 379} [\href{https://arxiv.org/abs/hep-th/9803001}{{\ttfamily
  hep-th/9803001}}].

\bibitem{Gomis:2006sb}
J.~Gomis and F.~Passerini, \emph{{Holographic Wilson Loops}},
  \href{https://doi.org/10.1088/1126-6708/2006/08/074}{\emph{JHEP} {\bfseries
  08} (2006) 074} [\href{https://arxiv.org/abs/hep-th/0604007}{{\ttfamily
  hep-th/0604007}}].

\bibitem{Polyakov:1980ca}
A.~M. Polyakov, \emph{{Gauge Fields as Rings of Glue}},
  \href{https://doi.org/10.1016/0550-3213(80)90507-6}{\emph{Nucl. Phys.}
  {\bfseries B164} (1980) 171}.

\bibitem{Gervais:1979fv}
J.-L. Gervais and A.~Neveu, \emph{{The Slope of the Leading Regge Trajectory in
  Quantum Chromodynamics}},
  \href{https://doi.org/10.1016/0550-3213(80)90397-1}{\emph{Nucl. Phys.}
  {\bfseries B163} (1980) 189}.

\bibitem{Dotsenko:1979wb}
V.~S. Dotsenko and S.~N. Vergeles, \emph{{Renormalizability of Phase Factors in
  the Nonabelian Gauge Theory}},
  \href{https://doi.org/10.1016/0550-3213(80)90103-0}{\emph{Nucl. Phys.}
  {\bfseries B169} (1980) 527}.

\bibitem{Fried:1958zz}
H.~M. Fried and D.~R. Yennie, \emph{{New Techniques in the Lamb Shift
  Calculation}}, \href{https://doi.org/10.1103/PhysRev.112.1391}{\emph{Phys.
  Rev.} {\bfseries 112} (1958) 1391}.

\bibitem{Yennie:1961ad}
D.~R. Yennie, S.~C. Frautschi and H.~Suura, \emph{{The infrared divergence
  phenomena and high-energy processes}},
  \href{https://doi.org/10.1016/0003-4916(61)90151-8}{\emph{Annals Phys.}
  {\bfseries 13} (1961) 379}.

\bibitem{Ivanov:1985bk}
S.~V. Ivanov and G.~P. Korchemsky, \emph{{SOME SUPPLEMENTS OF NONPERTURBATIVE
  GAUGES}}, \href{https://doi.org/10.1016/0370-2693(85)90584-2}{\emph{Phys.
  Lett.} {\bfseries 154B} (1985) 197}.

\bibitem{Ivanov:1985np}
S.~V. Ivanov, G.~P. Korchemsky and A.~V. Radyushkin, \emph{{Infrared
  Asymptotics of Perturbative {QCD}: Contour Gauges}}, {\emph{Yad. Fiz.}
  {\bfseries 44} (1986) 230}.

\bibitem{Korchemsky:1985xj}
G.~P. Korchemsky and A.~V. Radyushkin, \emph{{Loop Space Formalism and
  Renormalization Group for the Infrared Asymptotics of {QCD}}},
  \href{https://doi.org/10.1016/0370-2693(86)91439-5}{\emph{Phys. Lett.}
  {\bfseries B171} (1986) 459}.

\bibitem{Bourget:2018obm}
A.~Bourget, D.~Rodriguez-Gomez and J.~G. Russo, \emph{{A limit for large
  $R$-charge correlators in $\mathcal{N}=2$ theories}},
  \href{https://arxiv.org/abs/1803.00580}{{\ttfamily 1803.00580}}.

\bibitem{Korchemsky:1987wg}
G.~P. Korchemsky and A.~V. Radyushkin, \emph{{Renormalization of the Wilson
  Loops Beyond the Leading Order}},
  \href{https://doi.org/10.1016/0550-3213(87)90277-X}{\emph{Nucl. Phys.}
  {\bfseries B283} (1987) 342}.

\bibitem{Korchemsky:1991zp}
G.~P. Korchemsky and A.~V. Radyushkin, \emph{{Infrared factorization, Wilson
  lines and the heavy quark limit}},
  \href{https://doi.org/10.1016/0370-2693(92)90405-S}{\emph{Phys. Lett.}
  {\bfseries B279} (1992) 359}
  [\href{https://arxiv.org/abs/hep-ph/9203222}{{\ttfamily hep-ph/9203222}}].

\bibitem{Korchemskaya:1996je}
I.~A. Korchemskaya and G.~P. Korchemsky, \emph{{Evolution equation for gluon
  Regge trajectory}},
  \href{https://doi.org/10.1016/0370-2693(96)01016-7}{\emph{Phys. Lett.}
  {\bfseries B387} (1996) 346}
  [\href{https://arxiv.org/abs/hep-ph/9607229}{{\ttfamily hep-ph/9607229}}].

\bibitem{Alday:2007hr}
L.~F. Alday and J.~M. Maldacena, \emph{{Gluon scattering amplitudes at strong
  coupling}}, \href{https://doi.org/10.1088/1126-6708/2007/06/064}{\emph{JHEP}
  {\bfseries 06} (2007) 064} [\href{https://arxiv.org/abs/0705.0303}{{\ttfamily
  0705.0303}}].

\bibitem{Fiol:2011zg}
B.~Fiol and B.~Garolera, \emph{{Energy Loss of an Infinitely Massive
  Half-Bogomol'nyi-Prasad-Sommerfeld Particle by Radiation to All Orders in
  $1/N$}}, \href{https://doi.org/10.1103/PhysRevLett.107.151601}{\emph{Phys.
  Rev. Lett.} {\bfseries 107} (2011) 151601}
  [\href{https://arxiv.org/abs/1106.5418}{{\ttfamily 1106.5418}}].

\bibitem{Correa:2012at}
D.~Correa, J.~Henn, J.~Maldacena and A.~Sever, \emph{{An exact formula for the
  radiation of a moving quark in N=4 super Yang Mills}},
  \href{https://doi.org/10.1007/JHEP06(2012)048}{\emph{JHEP} {\bfseries 06}
  (2012) 048} [\href{https://arxiv.org/abs/1202.4455}{{\ttfamily 1202.4455}}].

\bibitem{Fiol:2012sg}
B.~Fiol, B.~Garolera and A.~Lewkowycz, \emph{{Exact results for static and
  radiative fields of a quark in N=4 super Yang-Mills}},
  \href{https://doi.org/10.1007/JHEP05(2012)093}{\emph{JHEP} {\bfseries 05}
  (2012) 093} [\href{https://arxiv.org/abs/1202.5292}{{\ttfamily 1202.5292}}].

\bibitem{Mueck:2010ja}
W.~Mueck, \emph{{The Polyakov Loop of Anti-symmetric Representations as a
  Quantum Impurity Model}}, \href{https://doi.org/10.1103/PhysRevD.83.066006,
  10.1103/PhysRevD.84.129903}{\emph{Phys. Rev.} {\bfseries D83} (2011) 066006}
  [\href{https://arxiv.org/abs/1012.1973}{{\ttfamily 1012.1973}}].

\bibitem{Beccaria:2017rbe}
M.~Beccaria, S.~Giombi and A.~Tseytlin, \emph{{Non-supersymmetric Wilson loop
  in $ \mathcal{N} $ = 4 SYM and defect 1d CFT}},
  \href{https://doi.org/10.1007/JHEP03(2018)131}{\emph{JHEP} {\bfseries 03}
  (2018) 131} [\href{https://arxiv.org/abs/1712.06874}{{\ttfamily
  1712.06874}}].

\bibitem{Beccaria:2018ocq}
M.~Beccaria and A.~A. Tseytlin, \emph{{On non-supersymmetric generalizations of
  the Wilson-Maldacena loops in N=4 SYM}},
  \href{https://arxiv.org/abs/1804.02179}{{\ttfamily 1804.02179}}.

\bibitem{Alday:2007he}
L.~F. Alday and J.~Maldacena, \emph{{Comments on gluon scattering amplitudes
  via AdS/CFT}},
  \href{https://doi.org/10.1088/1126-6708/2007/11/068}{\emph{JHEP} {\bfseries
  11} (2007) 068} [\href{https://arxiv.org/abs/0710.1060}{{\ttfamily
  0710.1060}}].

\bibitem{Polchinski:2011im}
J.~Polchinski and J.~Sully, \emph{{Wilson Loop Renormalization Group Flows}},
  \href{https://doi.org/10.1007/JHEP10(2011)059}{\emph{JHEP} {\bfseries 10}
  (2011) 059} [\href{https://arxiv.org/abs/1104.5077}{{\ttfamily 1104.5077}}].

\bibitem{Sachdev:1992fk}
S.~Sachdev and J.~Ye, \emph{{Gapless spin fluid ground state in a random,
  quantum Heisenberg magnet}},
  \href{https://doi.org/10.1103/PhysRevLett.70.3339}{\emph{Phys. Rev. Lett.}
  {\bfseries 70} (1993) 3339}
  [\href{https://arxiv.org/abs/cond-mat/9212030}{{\ttfamily
  cond-mat/9212030}}].

\bibitem{Kitaev:2015}
A.~Kitaev, \emph{{A Simple Model of Quantum Holography.. \ KITP Program
  ?Entanglement in Strongly-Correlated Quantum Matter,?}},
  \href{https://arxiv.org/abs/{unpublished; see
  http://online.kitp.ucsb.edu/online/entangled15/}}{{\ttfamily {unpublished;
  see http://online.kitp.ucsb.edu/online/entangled15/}}}.

\bibitem{Polchinski:2016xgd}
J.~Polchinski and V.~Rosenhaus, \emph{{The Spectrum in the Sachdev-Ye-Kitaev
  Model}}, \href{https://doi.org/10.1007/JHEP04(2016)001}{\emph{JHEP}
  {\bfseries 04} (2016) 001}
  [\href{https://arxiv.org/abs/1601.06768}{{\ttfamily 1601.06768}}].

\bibitem{Maldacena:2016hyu}
J.~Maldacena and D.~Stanford, \emph{{Remarks on the Sachdev-Ye-Kitaev model}},
  \href{https://doi.org/10.1103/PhysRevD.94.106002}{\emph{Phys. Rev.}
  {\bfseries D94} (2016) 106002}
  [\href{https://arxiv.org/abs/1604.07818}{{\ttfamily 1604.07818}}].

\bibitem{Almheiri:2014cka}
A.~Almheiri and J.~Polchinski, \emph{{Models of AdS$_{2}$ backreaction and
  holography}}, \href{https://doi.org/10.1007/JHEP11(2015)014}{\emph{JHEP}
  {\bfseries 11} (2015) 014} [\href{https://arxiv.org/abs/1402.6334}{{\ttfamily
  1402.6334}}].

\bibitem{Jensen:2016pah}
K.~Jensen, \emph{{Chaos in AdS$_2$ Holography}},
  \href{https://doi.org/10.1103/PhysRevLett.117.111601}{\emph{Phys. Rev. Lett.}
  {\bfseries 117} (2016) 111601}
  [\href{https://arxiv.org/abs/1605.06098}{{\ttfamily 1605.06098}}].

\bibitem{Maldacena:2016upp}
J.~Maldacena, D.~Stanford and Z.~Yang, \emph{{Conformal symmetry and its
  breaking in two dimensional Nearly Anti-de-Sitter space}},
  \href{https://doi.org/10.1093/ptep/ptw124}{\emph{PTEP} {\bfseries 2016}
  (2016) 12C104} [\href{https://arxiv.org/abs/1606.01857}{{\ttfamily
  1606.01857}}].

\bibitem{Sachdev:2010uj}
S.~Sachdev, \emph{{Strange metals and the AdS/CFT correspondence}},
  \href{https://doi.org/10.1088/1742-5468/2010/11/P11022}{\emph{J. Stat. Mech.}
  {\bfseries 1011} (2010) P11022}
  [\href{https://arxiv.org/abs/1010.0682}{{\ttfamily 1010.0682}}].

\bibitem{Erdmenger:2015xpq}
J.~Erdmenger, M.~Flory, C.~Hoyos, M.-N. Newrzella, A.~O'Bannon and J.~Wu,
  \emph{{Holographic impurities and Kondo effect}},
  \href{https://doi.org/10.1002/prop.201500079}{\emph{Fortsch. Phys.}
  {\bfseries 64} (2016) 322}
  [\href{https://arxiv.org/abs/1511.09362}{{\ttfamily 1511.09362}}].

\end{thebibliography}\endgroup

\end{document}